\documentclass[journal]{IEEEtran}
\usepackage{amsmath,amssymb,graphicx}
\usepackage{epsfig,color}

\begin{document}

\title{Linear Amplification of Optical Signal in Coupled Photonic Crystal Waveguides}

\author{Vakhtang Jandieri,~\IEEEmembership{Senior Member,~IEEE,}
        and~Ramaz~Khomeriki
\thanks{V. Jandieri is with Department of Computer and Electrical Engineering, Free
University of Tbilisi, 240 D. Agmashenebeli alley, 0159 Tbilisi,
Georgia, e-mail: jandieri@ee.knu.ac.kr. V. Jandieri is a Visiting
Scientist at Roma-Tre University, EMLAB Laboratory of
Electromagnetic Fields, Rome, Italy}
\thanks{R. Khomeriki is with Department of Physics, Tbilisi State University, 3
Chavchavadze, 0128 Tbilisi, Georgia}
}

\markboth{Journal of \LaTeX\ Class Files,~Vol.~11, No.~4, December~2012}%
{Shell \MakeLowercase{\textit{et al.}}: Bare Demo of IEEEtran.cls
for Journals}

\maketitle

\begin{abstract}
We introduce a weakly coupled photonic crystal waveguide as a
promising and realistic model for all-optical amplification. A
symmetric pillar type coupled photonic crystal waveguide consisting
of dielectric rods periodically distributed in a free space is
proposed as all-optical amplifier. Using the unique features of the
photonic crystals to control and guide the light, we have properly
chosen the frequency at which only one mode (odd mode) becomes the
propagating mode in the coupled photonic crystal waveguide, whereas
another mode (even mode) is completely reflected from the guiding
structure. Under this condition, the all-optical amplification is
fully realized. The amplification coefficient for the continuous
signal and the Gaussian pulse is calculated.
\end{abstract}


\section{Introduction}

\noindent {\it Photonic} bandgap crystals are artificial dielectric
or metallic structures in which any electromagnetic wave propagation
is forbidden within a fairly large frequency range. A photonic
crystal with a complete bandgap could be used to localize
electromagnetic waves to specific arrays, to guide and control the
propagation of the waves along certain directions at restricted
frequencies \cite{1,2}. Photonic crystal waveguide (PCW) can be made
by removing a row of either air columns or dielectric rods that
result in multimode guiding \cite{3}. PCWs devices are very
promising candidates for all optical telecommunication purposes due
to the absence of energy losses and their ability for strong light
confinement \cite{4,5}. If PCWs are placed in a close proximity, a
coupled PCW is formed and the optical power is efficiently
transferred from one PCW to another. Recently, the coupled PCWs have
received much attention because of their promising applications to
ultra-compact, miniaturized photonic devices such as filters,
switches, power dividers, and couplers.

In the manuscript we propose two weakly coupled PCWs as promising
and realistic experimental model for all-optical amplification. Our
idea of amplification of the optical signals in two weakly coupled
PCWs is based on the possibility of coexistence of two (even and
odd) fundamental modes, which have different propagation properties
at the same frequency. Particularly, if the light is launched into
two coupled PCWs at some particular frequency properly chosen, only
one mode becomes a propagating mode, whereas another mode is
completely reflected from the system and its carried power flux is
exactly zero. The operating setup is quite similar to the Y-junction
type waveguide analyzed in a detail in \cite{japan}. The novelty of
our work is that we give a deep physical insight into the
realization of the optical signal amplification in such a guiding
system.  It should be emphasized that all the scenarios of
all-optical amplification considered in the manuscript are
completely linear, thus there is no need for implementation of
nonlinear effects.

Very recently, in order to demonstrate the effect of all-optical
amplification, the authors have studied a simple structure, which is
composed of two coupled conventional dielectric waveguides separated
by silver metallic film and bounded by perfect electric conductors
(PEC) \cite{6}. Despite the fact that some promising results have
been obtained, due to sufficient losses in the metallic film, it is
very difficult to vividly demonstrate the effect of all-optical
amplifications and what it more important, it is difficult to
conduct the experiments. To avoid these difficulties, we propose
weakly coupled PCWs system as a realistic model for all-optical
amplification, because of their ability to realize the complete
confinement of slow light without metallic support. The authors
believe that these studies could open the possibility to design
novel devices with the wide practical application in all-optical
computing systems.

\begin{figure}[h] \begin{center}
\epsfig{file=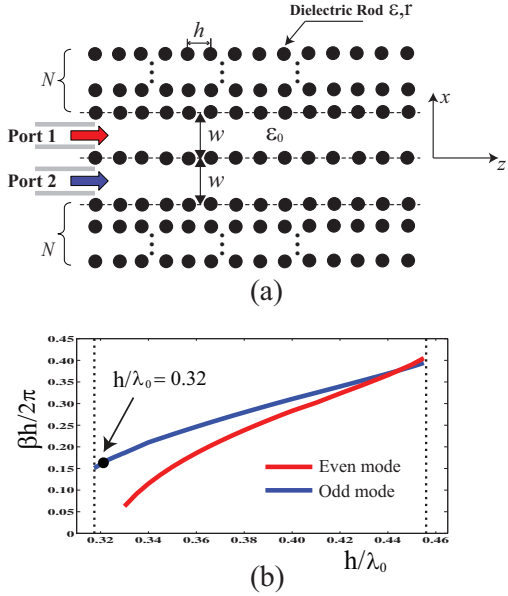,width=0.75\linewidth} \caption{(a) Schematic
view of the symmetric coupled two-parallel pillar-type PCWs
consisting of square lattice of dielectric rods periodically
distributed along z-axis in a free space:
$\varepsilon=11.56\varepsilon_0$, $r/h=0.175$ and $w/h=2.0$, where
$h$ is a period of the structure. Number of layers in both the upper
and lower halves of the photonic crystal is $N=10$ and the length of
the photonic crystal is $30h$. The signals are launched into the
coupled PCWs through Port 1 and Port 2 bounded by PEC walls marked
by gray lines. The structure is two-dimensional. (b) Dispersion
curves of even (red line) and odd (blue line) modes of the coupled
pillar-type PCWs shown in Fig. 1a. Bandgap region of the upper and
lower photonic crystals lies in the frequency range
$0.317<h/\lambda_0<0.458$ marked by dotted lines.} \label{lattice}
\end{center}
\end{figure}

\section{Numerical results and discussions}
The two coupled symmetric pillar type PCWs consisting of square
lattice of dielectric rods in free space is illustrated in Fig. 1a.
The guiding regions are bounded by the upper and lower photonic
crystals, which have a common period $h$ along the $z$-axis. The
parameters of the structure are chosen as $\varepsilon=11.56$,
$r/h=0.175$, $w/h=2.0$ and $N=10$ where $\varepsilon$ and $r$ are
the relative permittivity and radius of the dielectric circular
rods, $w$ is a width of the PCW and $N$ is the number of layers of
the upper and lower photonic crystals. The length of the photonic
crystals is $30h$. In the case of pillar type photonic crystal
waveguide, the modes are confined in the guiding region only due to
the existence of bandgap region, since the effective index of the
guiding region is smaller than that of the cladding region. Under
the adjusted parameters, the pillar type photonic crystal has a
photonic bandgap for $E$-polarized field $(E_y,H_x,H_z)$ in the
frequency range $0.317<h/\lambda_0< 0.458$, where $\lambda_0$ is a
wavelength in a free space. The dispersion diagrams of the coupled
PCWs (Fig. 1) are studied based on the coupled-mode formulation
\cite{7,8}, which we have recently proposed using the first-order
perturbation theory taking into account a weak coupling effect.
Based on the derived coupled-mode equations \cite{7,8} normalized
propagation constant $\beta h/2\pi$ of the pillar type coupled PCWs
(Fig.1) for the even mode and odd mode versus the normalized
frequency $h/\lambda_0$ is calculated and it is plotted in Fig. 1b
by red line and blue lines, respectively. It should be noted that no
other propagating modes exist in the frequency range
$0.317<h/\lambda_0< 0.458$. From the figure it follows that in the
lower frequency region $0.317<h/\lambda_0< 0.330$ the even mode
enters in a cutoff region and only odd mode is propagating along the
$z$-axis. From a viewpoint of the practical application of the
coupled PCWs structure as all-optical amplifier, we are interested
in the frequency region $0.317<h/\lambda_0< 0.330$, where only one
propagating mode exists. All the numerical analysis that follow are
conducted at the fixed normalized frequency $h/\lambda_0=0.320$.
\begin{figure}[t] \begin{center}
\epsfig{file=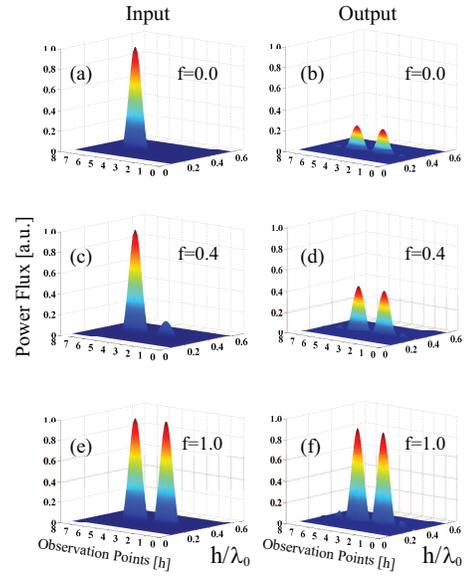,width=0.7\linewidth} \caption{Power flux of the
input and output signals at $f=0$, $f=0.4$ and $f=1.0$, where $f$ is
an amplitude of a signal launched into Port 2. Signal with unit
amplitude is injected in Port 1. The output signal is registered by
a detector located at a distance $z=10h$ inside the coupled PCWs.
The length of the detector is $8h$ and it is parallel along the
$x$-axis.} \label{pow}
\end{center}
\end{figure}

To validate the use of the coupled PCWs as all-optical amplifier,
firstly, we analyze the power flux of a continuous signal launched
into the coupled PCWs (Fig.1) using the FDTD method \cite{9,10}. We
have chosen the discretization step size as $\Delta x=\Delta
z=h/20$, where $h$ is the lattice period. The discretized time step
is $\Delta t=1.12\cdot 10^{-10}$, by which the stability condition
of FDTD method is satisfied. The analysis region is surrounded by
Berenger's PML with thickness of 20 cells in all of the surrounding
walls. The initial excitations are injected into the coupled PCWs at
$z=0$ through Port 1 and Port 2. Input Ports are bounded by PEC
walls marked by gray lines in Fig. 1. The output signal is
registered by the detector placed at a distance $z=10h$ inside the
coupled PCWs. The length of the detector is $\ell=8h$ and it is
parallel along the $x$-axis. To demonstrate the effect of
all-optical amplification, firstly we inject a signal with unit
amplitude in Port 1, whereas no signal $f=0$ is launched in Port 2.
The power flux $P_{f=0}^{Input}$ of the total input signal is
illustrated in Fig. 2a. In this case the total input could be
represented as a sum of even and odd modes and since the coupled
PCWs support only the odd mode at $h/\lambda_0=0.320$ (Fig. 1b), the
even mode is completely reflected and the total output power
$P_{f=0}^{Output}$ is half of the input power as shown in Fig. 2b.
Next, we inject a continuous signal in Port 2 with an amplitude
$f=0.4$. The total input power flux $P_{f=0.4}^{Input}$ is
illustrated in Fig. 2c. It is important to mention that in order to
guide an input signal into the coupled PCWs, which supports only one
odd mode, the injected signals in Port 1 and Port 2 should be with
opposite phases. The output power flux $P_{f=0.4}^{Output}$ is
presented in Fig. 2d. The amplification coefficient is calculated as
\begin{equation}
R=\frac{\int_\ell P_{f=0.4}^{Output}dx-\int_\ell
P_{f=0}^{Output}dx}{\int_\ell P_{f=0.4}^{Input}dx-\int_\ell
P_{f=0}^{Input}dx} \label{eq1}
\end{equation} and it is equal to $R=2.82$. It is worth mentioning that in case of
weakly coupled PCWs the result for the amplification coefficient is
very close to the result obtained using the rigorous theoretical
analysis \cite{6}
\begin{equation}
R=0.5+(1/f).
\end{equation}
Additionally, our numerical analysis have vividly demonstrated that
the amplification coefficient is substantially improved in
comparison to the results for the conventional dielectric waveguides
separated by silver metallic film \cite{6}. It is obvious that an
increase of the amplification coefficient is caused by the absence
of energy losses in pillar type coupled PCWs. Finally, we further
increase the amplitude of the injected signal in input Port 2 up to
$f=1.0$ and the power flux of the input $P_{f=1}^{Input}$  and
output $P_{f=1}^{Output}$ signals are illustrated in Fig. 2e and
Fig. 2f, respectively. The amplification coefficient is $R=1.45$.
The results for the amplification coefficients $R$ at different
amplitudes $f$ are presented in some detail in Table 1. From the
results it follows that the amplification coefficient is inversely
proportional to the amplitude $f$ of the input signal.

\begin{figure}[t] \begin{center} \hspace{-2.5cm}
\epsfig{file=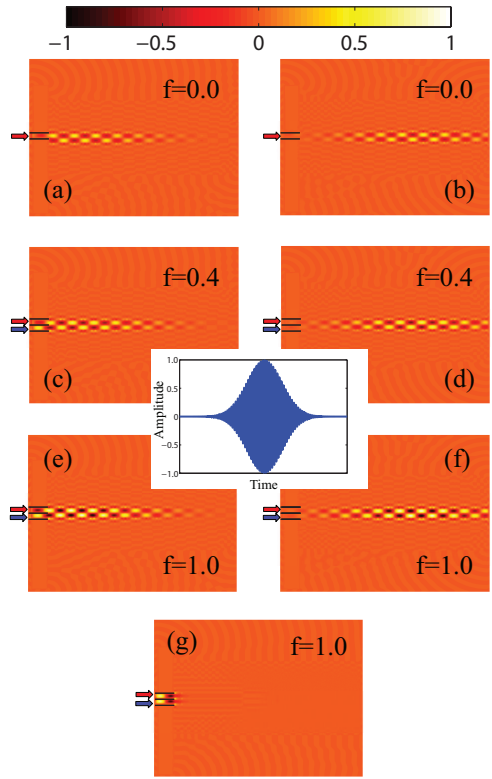,width=1.\linewidth} \caption{Near field
distribution of Gaussian pulse propagating along the z-axis at $f=0$
(a),(b), $f=0.4$ (c),(d) and $f=1.0$ (e),(f), when the input signals
with opposite phases (odd mode) are launched into Port 1 and Port 2.
In (g) the input signals with the same phases (even mode) are
injected in Port 1 and Port 2 at $f=1.0$.} \label{pulse}
\end{center}
\end{figure}

Next, in Fig. 3 we are analyzing the near field distributions of the
Gaussian pulse propagating in the weakly coupled PCWs. The time
waveform of the Gaussian pulse is centered at the time step 4500 and
the full width at half maximum (FWHM) of the pulse is 1500 time
steps. Near field distributions at $f=0$, $f=0.4$ and $f=1.0$ at
different moments of time are presented in Fig. 3a - Fig. 3f,
respectively. The amplification coefficient of Gaussian pulse is
calculated and the results are presented in Table 1. Similarly to
the previous case of continuous signal, a good agreement between our
results and the case considered in \cite{6}. In comparison, the near
field distribution of the Gaussian pulse injected in Port 1 and Port
2 with the same phase (even mode) is illustrated in Fig. 3g. The
pulse in completely reflected from the coupled PCWs and there is no
power carried inside the coupled PCWs. This is due to the fact that
the coupled pillar type PCWs do not support the even mode at the
normalized frequency $h/\lambda_0$ (Fig. 1b). The coupled PCWs
behave as a phase filter providing maximum output power for the
signals with the opposite phases (Fig. 3a - Fig. 3f) and zero power
for the signal with the same phases (see Fig. 3g).

\begin{table}[h]
\caption{\label{12} \bf Amplification coefficient $R$ at different
amplitudes $f$ of the input signal launched into two coupled PCW-s
(Fig.1a). The results are compared with those obtained by the
rigorous analysis \cite{6}.}
\centering
\begin{tabular}{|c|c|c|c|}
\hline
$f$ & Continuous & Gaussian & Rigorous Analysis \cite{6} \\
 & Signal & Pulse & $R=0.5+(1/f)$ \\
\hline
0.2 & 5.30 & 4.90 & 5.50 \\
\hline
0.4 & 2.82 & 2.70 & 3.00 \\
\hline
0.5 & 2.38 & 2.29 & 2.50 \\
\hline
0.8 & 1.68 & 1.62 & 1.75 \\
\hline
1.0 & 1.45 & 1.41 & 1.50 \\
\hline
\end{tabular}
\end{table}

\section{Conclusion}
In this manuscript, for the first time, we have proposed a weakly
coupled photonic crystals waveguide as a realistic model for
all-optical amplifier.
It should be noted that the amplification effect of the optical
signal proposed in the manuscript is quite different from that in
homodyne receiver scheme, where the signal intensity is amplified at
the receiving area only, whereas the total signal is not amplified.
The authors believe that one of the promising practical applications
of the proposed structure could be optical transistors. Being able
to precisely control the intensity of one light beam using another
could be crucial to the development of optical transistors capable
of performing complex light-controlling in all-optical circuits.


\end{document}